\documentstyle[aps,prl,twocolumn,epsfig]{revtex}
\begin{document}
\twocolumn[\hsize\textwidth\columnwidth\hsize\csname @twocolumnfalse\endcsname
\draft

\title{Pseudogaps and their Interplay with Magnetic Excitations\\ in the
  doped 2D Hubbard Model}
\author{R. Preuss \cite{roland}, W. Hanke, C. Gr\"{o}ber and H. G. Evertz}
\address{Institut f\"{u}r Theoretische Physik, Am Hubland,
D-97074 W\"{u}rzburg, Federal Republic of Germany\\
Email: hanke@physik.uni-wuerzburg.de}
\date{\today}
\maketitle

\begin{abstract}
On the basis of Quantum Monte Carlo simulations of the two-dimensional
Hubbard model which cover the doping range from the under- to the 
over-doped regime, we find that the single-particle spectral weight 
$A (\vec k,\omega)$ qualitatively reproduces both the momentum
($d_{x^2-y^2}$--symmetry) and doping dependence of the pseudogap
as found in photoemission experiments. The drastic doping dependence
of the spin response $\chi_{s} (\vec q,\omega)$ which is sharp 
in both $\vec q \ (\approx(\pi,\pi))$ and $\omega$ in the under-doped 
regime but broad and structureless otherwise, identifies remnants of 
the antiferromagnetic order as the driving mechanism behind the 
pseudogap and its evolution with doping.
\end{abstract}
]

Exciting progress in the microscopic understanding of high-$T_{C}$
superconductors has recently come from the observation of a 
normal-state pseudogap of order of the exchange energy 
$J$ \cite{01} and a lower energy excitation gap of the 
order of the superconducting gap \cite{01,02,03,04,08}.
Angle-resolved photoemission spectroscopy (ARPES) demonstrated 
that both high- and low-energy gaps are consistent 
with $d_{x^2-y^2}$--symmetry \cite{01,02,03,04}. 
In addition, both gaps have a more or less identical doping 
dependence, which may be a key observation to unlocking the 
mystery of the cuprates: just below optimal hole concentration,
the centroids in the spectral weight $A (\vec k,\omega)$ near 
$(\pi,0)$ move to higher binding energy, and a portion of 
the large Fermi surface seems to disappear. These findings have
been interpreted as the opening of a pseudogap with maximal energy $J
\sim 200 meV$ near $(\pi,0)$ \cite{01}. Simultaneously, a normal-state
gap with energy $\sim 20 meV$, inferred from the leading edge in 
$A (\vec k,\omega)$, opens up in this under-doped regime. Both of 
these gaps vanish in the over-doped regime, and the superconducting 
gap also rapidly decreases \cite{08}. This empirical correlation 
between the disappearance of the order $J$ pseudogap and the decrease 
of superconducting pairing strength suggests that the high-energy 
features at $(\pi,0)$ are closely related to the pairing interaction 
\cite{09}.

In this letter, we address the microscopic mechanism behind the
opening of this pseudogap and its evolution from under-doped to 
over-doped regimes. We present Quantum Monte Carlo (QMC) simulation 
results on the two-dimensional Hubbard model with on-site 
interaction $U=8t$ which demonstrate that the single-particle 
spectral weight $A (\vec k,\omega)$ reproduces the most salient 
ARPES features as function of doping. In particular, the QMC data 
reproduce the momentum ($d_{x^2-y^2}$--symmetry) and doping dependence
of the pseudogap. 

Earlier finite-temperature QMC work on the Hubbard model \cite{16,16a}
has produced results showing a quasi-particle-like band with a dispersion 
below the Fermi level that is essentially unaffected by doping.
On the other hand, groundstate exact diagonalizations \cite{25,25a}
of the two-dimensional $t-J$ model for small clusters around optimal
doping find a signal in the spectral weight corresponding to an
insulator-like ``shadow'' structure; at larger doping this signal
vanishes. We present in this work first data of the spectral weight 
$A (\vec k, \omega)$ obtained by Maximum-Entropy techniques for 
previously inaccessible temperatures ($T=0.25t$), showing that 
if the temperature in the dynamical QMC simulation is lowered 
below a threshold temperature $T^{\ast} \simeq 0.3t$ in the 
under-doped regime the quasi-particle band is substantially 
deformed, resulting in the opening of a pseudogap. 

Our results provide numerical evidence of earlier theoretical 
conjectures that the deformation and the pseudogap are intimately 
related to antiferromagnetic spin fluctuations \cite{10,11,12}. 
This is demonstrated by studies of the two-particle excitation 
spectrum over a wide range of dopings from under-doped to over-doped. 
They reveal that it is the strongly doping-dependent spin response 
and not the essentially doping-independent charge response which 
tracks the doping dependence of the pseudogap. The spin response, $\chi_{s} 
(\vec q,\omega)$, displays a sharp structure in $\mbox{Im} \ \chi_{s}$
at wave-vector $\vec q \cong \vec Q \equiv (\pi,\pi)$ and at energy
$\omega (\vec Q) = \omega^{\ast}$. Not only near half-filling but also
up to optimal doping the spectral weight is distributed around an
energy dispersion which still closely follows the spin-wave dispersion
$\omega (\vec q) \sim \omega^{SDW} (\vec q)$, where $\omega^{SDW}$ 
is calculated within the spin-density-wave approximation. For 
temperatures $T < T^{\ast}$, the antiferromagnetic correlation 
length $\xi (T)$ becomes larger than the lattice spacing $a_{0}$ 
and, as a consequence, the quasi-particle is strongly dressed by 
spin fluctuations. As soon as the system enters the over-doped regime, 
the spin response is no longer sharply peaked near $\vec Q=(\pi,\pi)$ 
and $\omega^{\ast}$. It spreads in energy by an order of magnitude 
(the scale changes from $J$ to $E_{kin} \sim 8t$), and is accompanied 
by a similar change of the scale of the bandwidth for single-particle 
excitations. Thus it is this unique doping dependence of the magnetic 
excitations which establish the pseudogap as being due to 
antiferromagnetic spin fluctuations.

The single-band two-dimensional Hubbard model has the standard Hamiltonian 
\begin{eqnarray}
  H = -t \sum_{\langle ij \rangle , \sigma} \left(
    c^{\dagger}_{i,\sigma} c_{j, \sigma} + h.c. \right) 
  + U \sum_{i} n_{i \uparrow} n_{i \downarrow} \nonumber \\
  - \mu \sum_{i} \left( n_{i \uparrow} + n_{i \downarrow} 
  \right) \label{hubham}
\end{eqnarray}  
on a square lattice, where $t$ is the nearest-neighbor hopping,
$c_{i,\sigma}$ destroys a particle of spin $\sigma$ on site $i$, and
$n_{i \sigma} = c^{\dagger}_{i,\sigma} c_{i,\sigma}$. The chemical
potential $\mu$ sets the filling $\langle n \rangle = \langle n_{i
  \uparrow} + n_{i \downarrow} \rangle$. The spectral weight 
$A (\vec k,\omega)$ is inferred from high-quality \cite{15a} QMC data by 
applying state-of-the-art ``Maximum-Entropy'' techniques \cite{15}. 
This method has previously been used to resolve a quasi-particle-like 
dispersive band of width $J$ at half-filling and to follow 
its evolution from the insulator to the metal \cite{16}.

\begin{figure}

  \vspace*{-0.25cm}

  \epsfig{file=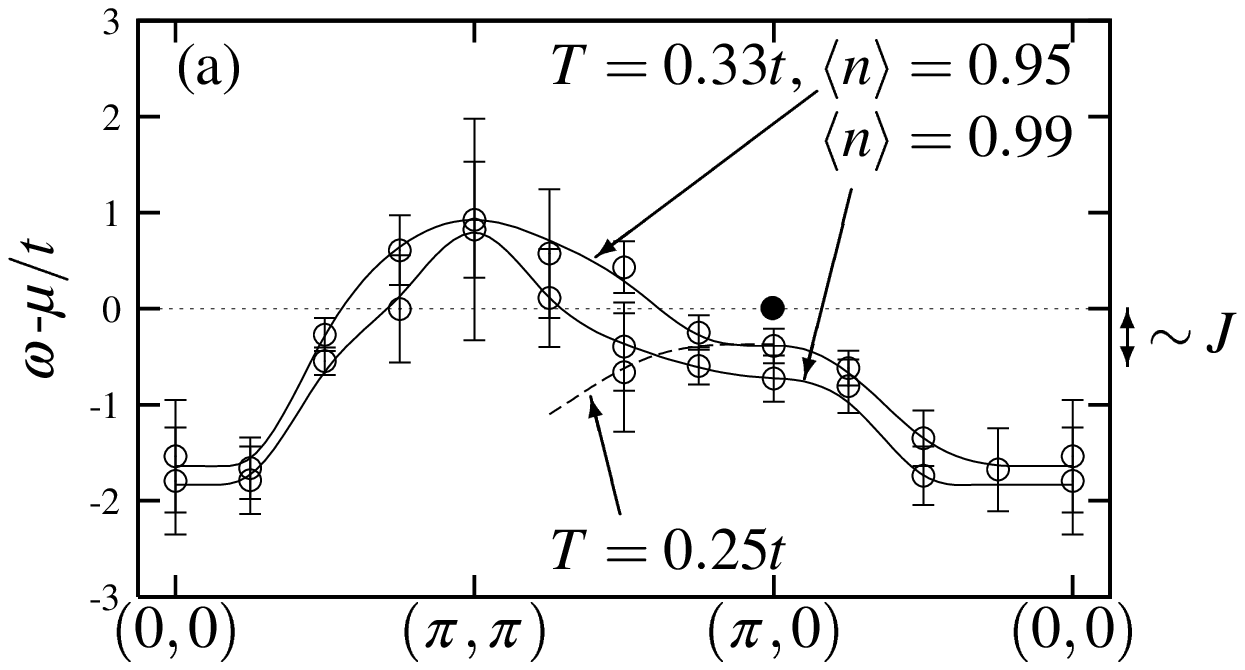,width=8.50cm}

  \vspace*{0.25cm}
  \hspace*{-0.40cm}
  \epsfig{file=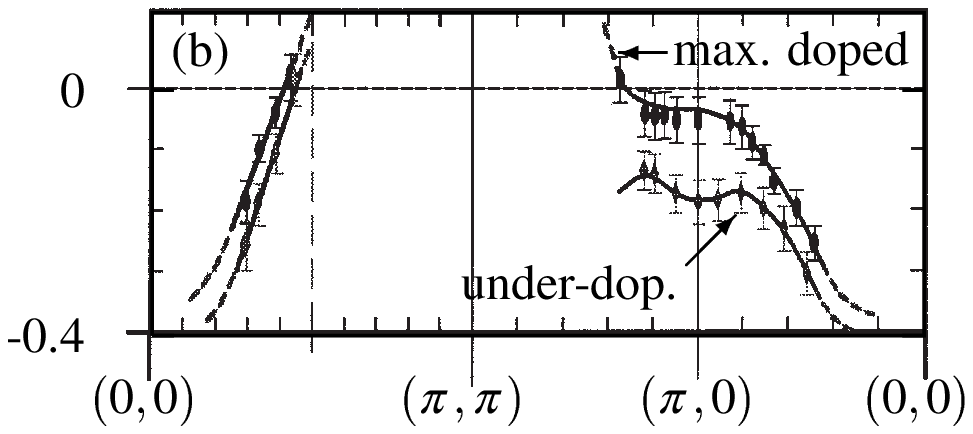,width=8.25cm}\\[-0.20cm]

  \caption{The dispersion of the peaks in the single-particle spectral
    weight from (a) QMC simulations of the Hubbard model at densities 
    ranging from the under-doped to the maximally doped regime where 
    peaks in $A(\vec k,\omega)$ are represented by error bars and (b) 
    peak centroids in $A(\vec k,\omega)$ from under-doped and
    maximally doped ARPES experiments after Ref. [1].}
  \label{f01}
\end{figure}

In Fig. \ref{f01}(a) we compare the QMC results for the single-particle 
spectral weight $A (\vec k,\omega)$ for the under-doped ($1\%$ and $5\%$) 
and maximally doped ($13\%$) regimes with the ARPES data from Ref. 
\cite{01} in Fig. \ref{f01}(b). The theoretical results were obtained 
for temperatures of $T=0.33t$ and $T=0.25t$, on-site Coulomb repulsion
$U=8t$ and $8 \times 8$ lattices. To facilitate detailed comparison
with Fig. \ref{f01}(b) of D. S. Marshall, Z.-X. Shen {\it{et al.}} 
\cite{01}, we use in Fig. \ref{f01}(a) the same $\omega$ vs $\vec k$ 
``band structure'' plot with all spectra referring to the same
chemical potential $\mu \ (\omega=0)$.

In accordance with the ARPES data, we find a dramatic change in the 
electronic structure near $(\pi,0)$ with doping. In the under-doped
($1\%$ and $5\%$) regime, the features near $(\pi,0)$ move to lower
binding energy as doping is increased. At about $13\%$ doping
\cite{16a} (full circle in Fig.\ \ref{f01}(a)) the pseudogap vanishes.

In addition, a portion of the large Fermi surface, which is closed 
around $(\pi,\pi)$ at $T=0.33t$ (i.e. has Fermi level crossings 
between $(\pi,0)$ and $(\pi,\pi)$ as well as between $(\pi,\pi)$ 
and $(0,0)$), seems to disappear at the lower temperature,
$T=0.25t$. This is indicated in Fig. \ref{f01}(a) by the downturn of
the quasi-particle-like band between $(\pi,0)$ and $(\pi,\pi)$. 
In the experiment, this behavior in the under-doped regime has 
been interpreted as the opening of a pseudogap in the underlying 
Fermi surface near the $(\pi,0)$ to $(\pi,\pi)$ line \cite{01}.
The momentum dependence of the pseudogap is also in agreement with 
experiment in that its largest value, which is of the order of the 
exchange coupling $J \sim 4t^2/U$, occurs near $(\pi,0)$ and in that 
it essentially vanishes along the $(0,0)$ to $(\pi,\pi)$ line. This 
behavior is consistent with $d_{x^2-y^2}$--symmetry. At $13\%$ doping
we obtain a Fermi surface crossing near $(\pi,0)$ (full circle in
Fig.\ \ref{f01}(a)) independent of the two chosen temperatures. Thus, 
at optimal hole doping the Fermi surface is large, consistent with 
both ARPES \cite{01,02,03,04} and earlier QMC calculations \cite{16a}.

\begin{figure}

  \vspace*{-0.25cm}

  \epsfig{file=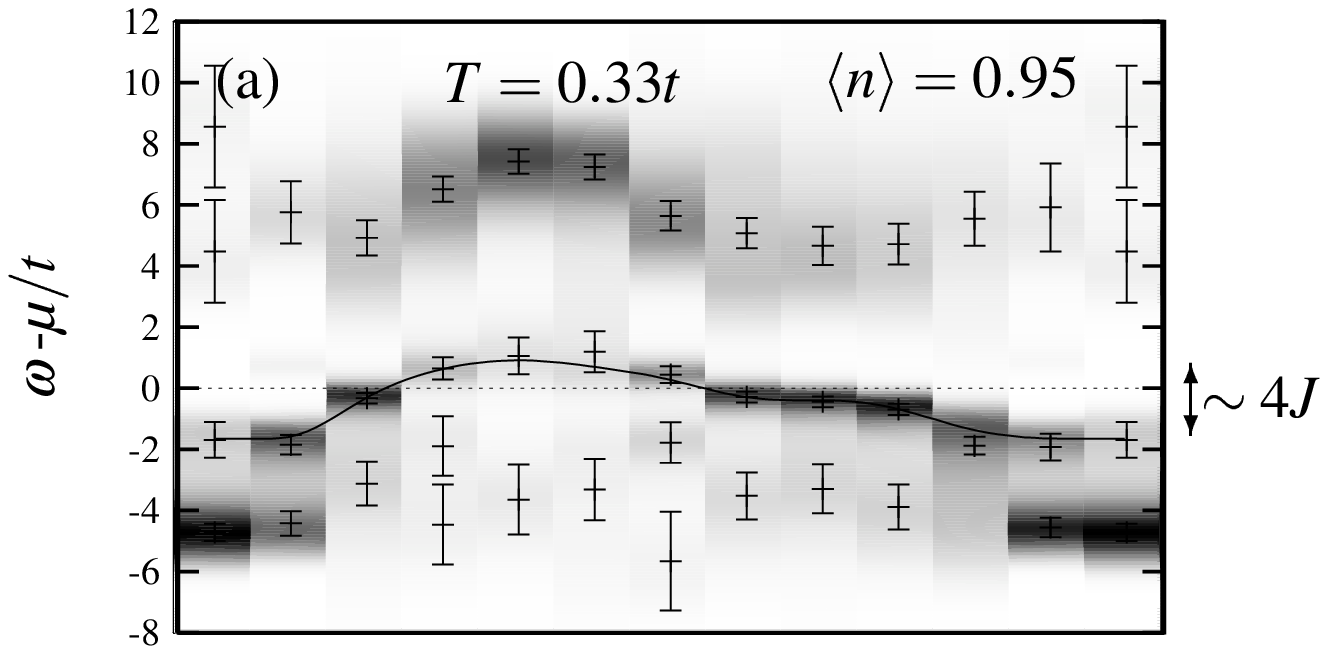,width=8.50cm}\\[-0.25cm]

  \epsfig{file=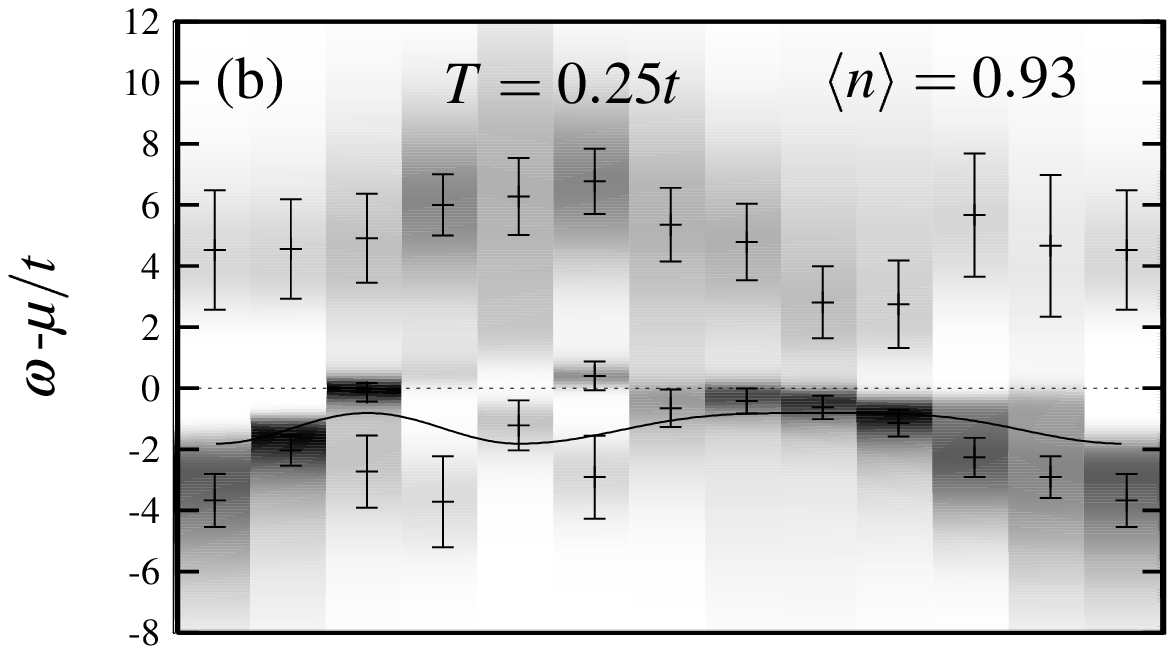,width=6.95cm}\\[-0.25cm]

  \epsfig{file=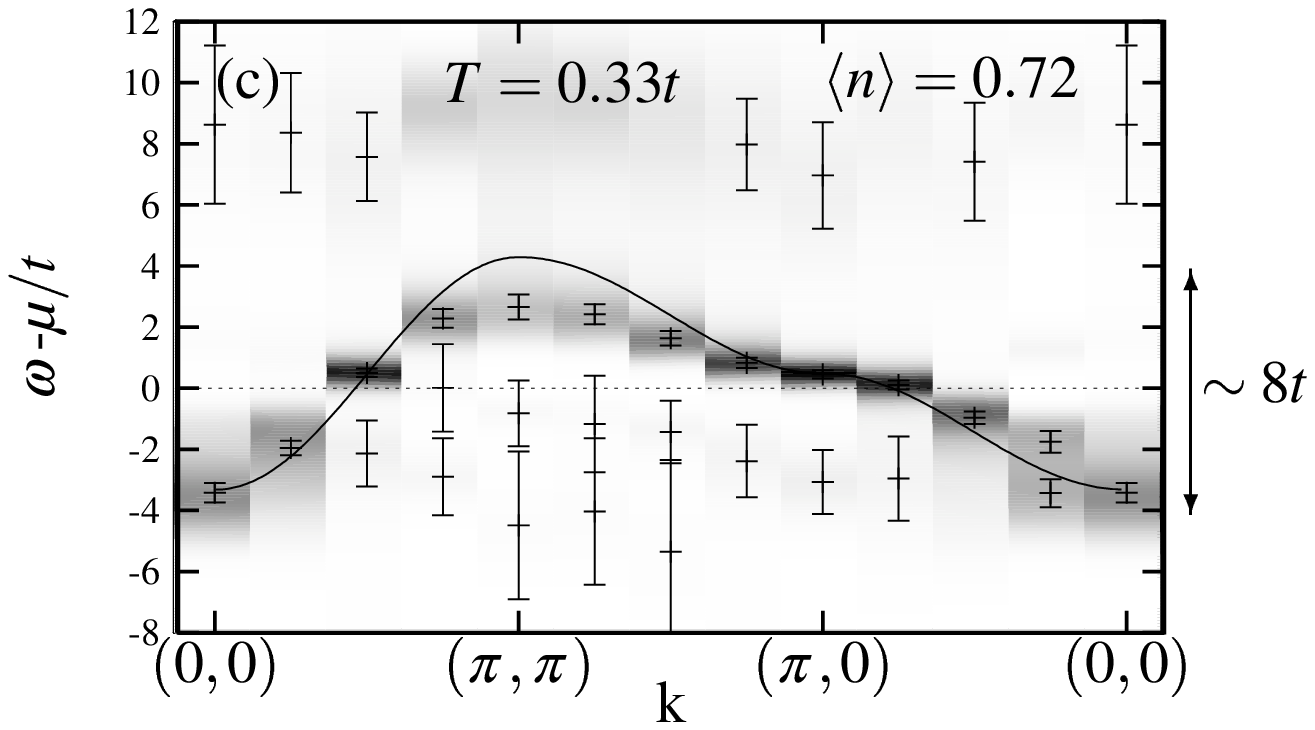,width=8.50cm}\\[-0.25cm]

  \caption{The QMC spectral weight, $A (\vec k,\omega)$, of the 
    $8 \times 8$ Hubbard model with $U=8t$ as a function of $\omega$
    and $\vec k$: (a) at temperature $T=0.33t$ and $5\%$ doping 
    ($\langle n \rangle = 0.95$), (b) at $T=0.25t$ and $7\%$ 
    doping ($\langle n \rangle = 0.93$) and, (c) at $T=0.33t$ 
    and $28\%$ doping. Dark (white) areas correspond to large 
    (small) spectral weight. The full lines in (a) and (c) are 
    tight-binding fits to the QMC data, the full line in (b) 
    denotes the $U=8t$ QMC result for $\langle n \rangle = 1.0$, 
    i.e. the insulating case.}
  \label{f02}
\end{figure}

In the QMC simulation, the opening of a pseudogap in the under-doped 
regime shows up below a ``crossover'' temperature $T^{\ast} \simeq
0.3t$. To illustrate this crossover, we consider in Fig. \ref{f02}(a) -
\ref{f02}(b) the role played by temperature and its influence on the 
spectral weight distribution in more detail. We observe two general
features in both spectra: a (several $t$) broad ``incoherent background'' 
both below and above the Fermi surface and a dispersing structure with a 
smaller width of order of a few $J$ around $\omega=\mu$. In our earlier 
work \cite{16} in which the lowest temperature accessible was $T=0.33t$, 
this dispersing quasi-particle-like band was shown to have its maximum at 
$(\pi,\pi)$. In this case, which is reproduced in Fig. \ref{f02}(a), there 
is a ``large'' Fermi surface centered around $(\pi,\pi)$. However, by 
lowering the temperature to $T=0.25t$, the structure forming the maximum 
at $(\pi,\pi)$ looses weight and the new valence-band maximum seems to
be shifted to $(\pi/2,\pi/2)$ or $(\pi,0)$ (Fig. \ref{f02}(b)) \cite{18a}. 
In agreement with the ARPES data in Fig. \ref{f01}(b), we observe the 
downturn (``shadow'' structure) as well as a drastic (by about a
factor of 10) spectral weight loss when following the quasi-particle 
band from $(\pi,0)$ to $(\pi,\pi)$. It is well known \cite{22a} that 
inclusion of higher-neighbor ($t'$, etc.) interactions lifts the 
degeneracy of the points $(\pi/2,\pi/2)$ and $(\pi,0)$ at 
half-filling and pushes the $(\pi/2,\pi/2)$ point up in energy. 
This opens up the possibility of a small Fermi surface (``hole 
pockets'') around $(\pi/2,\pi/2)$ \cite{25a,22}. The full line 
in Fig. \ref{f02}(b), which plots our results for the same 
temperature for  the insulating case, $\langle n \rangle = 1.0$, 
indicates the similarities between the $7\%$ doped situation and 
the antiferromagnetic band structure at half-filling.

\begin{figure}

  \vspace*{-0.25cm}

  \epsfig{file=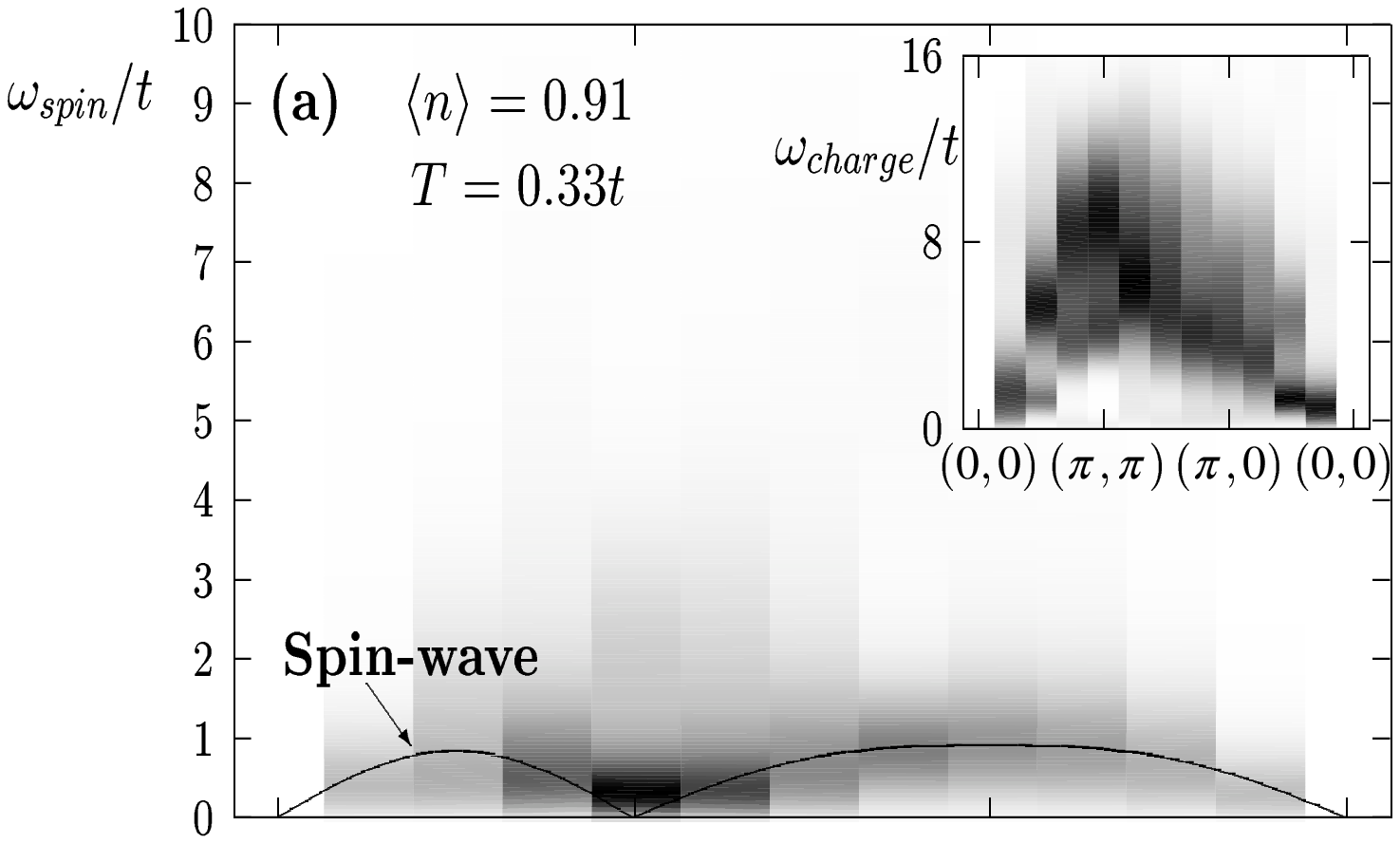,width=8.25cm}

  \epsfig{file=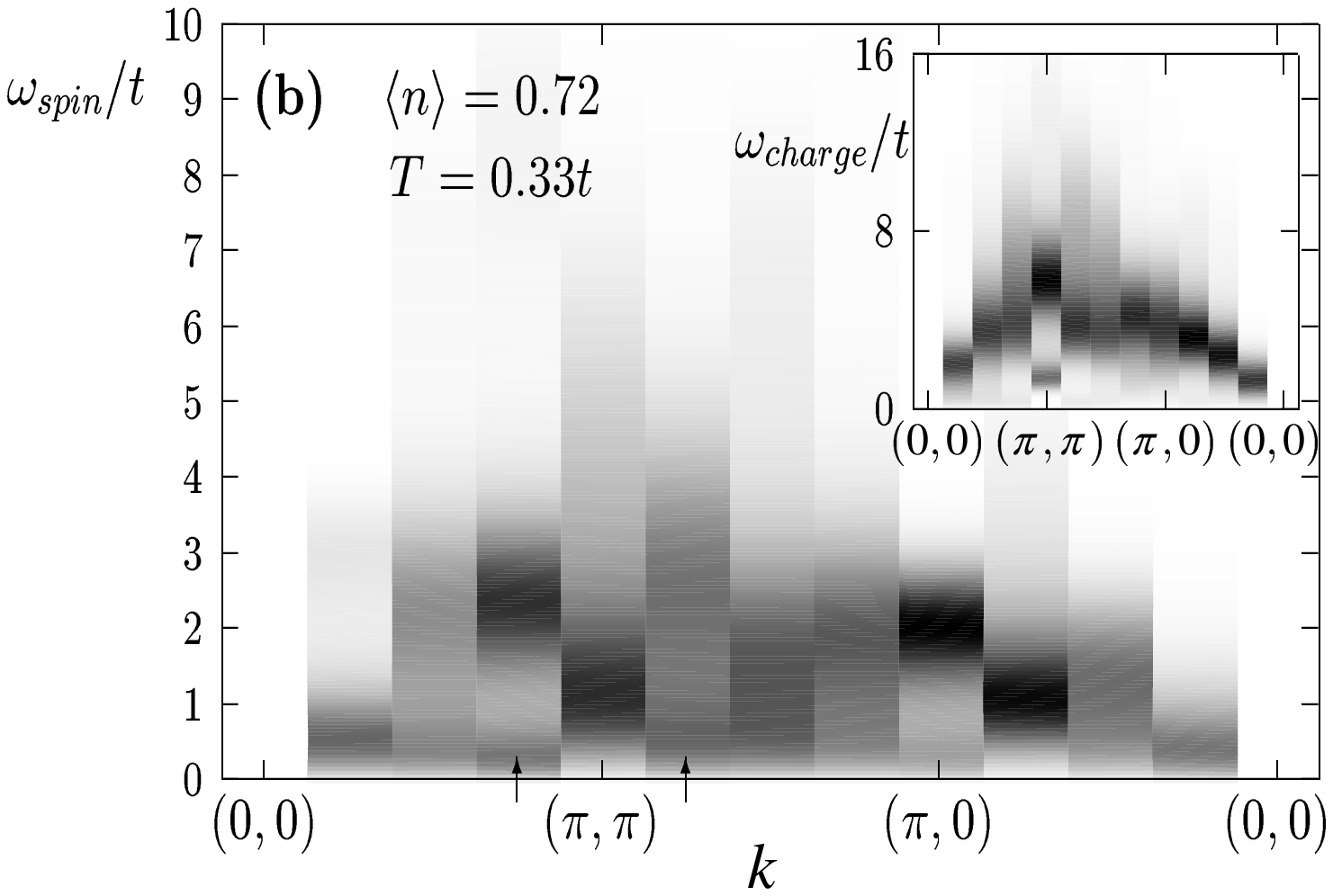,width=8.25cm}

  \vspace*{0.00cm}

  \caption{Two-particle spin ($\omega_{spin}$) and charge
    ($\omega_{charge}$ in the inset) excitations as function of $\vec
    k$, (a) under-doped ($\langle n \rangle =0.91$), (b) over-doped 
    ($\langle n \rangle =0.72$). Note the drastic change in the 
    spin response as function of doping.}
  \label{f03}
\end{figure}

The physical origin of a possible manifestation of insulating band features 
in the under-doped regime seems to be that around the temperature $T^{\ast} 
\simeq 0.3t$ the antiferromagnetic correlation length $\xi$ becomes 
larger than the lattice spacing $a_{0}$, i.e. for $T=0.25t$ we find 
$\xi \simeq 1.2 > 1$, whereas for $T=0.33t$ we find 
$\xi \simeq 0.5 < 1$ \cite{21}. The decisive role played 
by antiferromagnetic spin fluctuations for the pseudogap and the 
under-doped band structure is substantiated by the behavior 
of $\chi_{s,c} (\vec q,\omega)$, shown in Fig. \ref{f03}. 

Consider first Fig. \ref{f03}(a), which exhibits the spin excitation energy
as function of momentum for the $8 \times 8$ Hubbard model with
$U=8t$ and $T=0.33t$ at a doping of $9\%$ away from half-filling. Here, 
dark (white) areas again correspond to a large (small) spectral weight for 
spin excitations and the full line gives the spin-wave dispersion,
i.e. $\omega^{SDW} (\vec q) = 2J ( 1 - ( \varepsilon_{\vec q} /4t )^{2}
)^{1/2}$, where $\varepsilon_{\vec q}$ denotes the tight-binding energy
$\varepsilon_{\vec q} = -2t ( \cos (q_{x}) + \cos (q_{y}) )$.
This figure reveals that in the under-doped regime up to maximal doping 
($13\%$) the spin excitations have an energy dispersion that
still closely follows the antiferromagnetic spin-wave dispersion.
The main weight is confined to $\vec Q = (\pi,\pi)$ with a small
spread $\Delta \vec Q$ and to a small but finite energy $\omega(\vec Q) = 
\omega^{\ast}$. This sets an energy scale $J$ for spin excitations.
This result is in accordance with recent neutron-scattering 
data \cite{13} and also with other experiments that have directly 
shown that propagating spin-waves with energies of order $J$
still exist even at optimal doping \cite{14}. When the temperature 
is lowered below $T^{\ast} \simeq 0.3t$, $\Delta \vec Q$ slightly 
decreases and the antiferromagnetic correlation length $\xi(T)$ 
($\sim 1/\Delta Q$) becomes larger than the lattice spacing. As 
a consequence, the single-particle hopping is now strongly 
renormalized by the short-range antiferromagnetic order resulting 
in a bandwidth (Fig. \ref{f02}(b)) also of order of (a few) $J$. 
This renormalization is strongest at $(\pi,0)$ and, thus, directly 
responsible for the pseudogap: because of the ``matching''condition, 
$(\pi,0) \pm \vec Q = (0,\pi)$, the bare photohole created by ARPES 
for $\vec k=(\pi,0)$ couples strongly to two-particle excitations 
whose spectral function peaks near momentum $\vec Q=(\pi,\pi)$
\cite{09}. Using a standard diagrammatic evaluation of the self-energy
like the FLEX summation \cite{22b}, and approximating the peak in
$\mbox{Im} \ \chi_{s} (\vec q, \omega)$ by $\mbox{Im} \ \chi_{s} \sim 
\delta (\vec q - \vec Q) [ - \delta (\omega - \omega^{\ast} ) + 
\delta (\omega + \omega^{\ast} ) ]$, one obtains for $\omega < 
- \omega^{\ast} < 0$ the self-consistency equation
\begin{eqnarray}
  A(\vec k, \omega) \simeq \frac{U^{2} A(\vec k - \vec Q, \omega +
  \omega^{\ast})}{\left[ \left(\omega - \varepsilon_{\vec k} - 
  \mbox{Re} \ \Sigma (\vec k, \omega) \right)^{2} \! + \! \left. 
  \mbox{Im} \ \Sigma (\vec k, \omega) \right.^{2} \right]} , 
  \label{selfcons}
\end{eqnarray}
since $\mbox{Im} \Sigma (\vec k,\omega) \sim -U^{2} A(\vec k - \vec Q,
\omega + \omega^{\ast})$. As a consequence of this self-consistency 
requirement, multiple, repeated spin-wave excitations (``shake ups'') 
accompany the bare carrier motion, lead to incoherent contributions to
the electronic spectrum at $(\pi,0)$ and spread spectral weight to
lower energies. In the over-doped regime, shown in Fig. \ref{f03}(b), 
the sharpness both in $\vec q$ and $\omega$ in $\mbox{Im} \ \chi_{s} 
(\vec q,\omega)$, and thus the ``matching''condition is lost completely 
and the coupling to the bare photohole is weak and broad in energy. The 
crossover to a new energy scale ($t$ rather than $J$) is also reflected 
in a corresponding crossover in the single-particle bandwidth ($\sim 8t$) 
in Fig. \ref{f02}(c). In contrast to the spin excitations, the charge 
excitations are already broad and structureless in the under-doped 
regime, with the energy spreading essentially over the non-interacting
bandwidth $\sim 8t$ (inset in Fig. \ref{f03}(a)) \cite{24}.

In this letter, we have addressed the microscopic mechanism behind the
opening of a pseudogap and its evolution from under-doped to over-doped 
regimes. Our key results are (i) that the dynamical QMC results for
the two-dimensional Hubbard model reproduce both the momentum
(i.e. the $d_{x^2-y^2}$--symmetry) and doping dependence of the 
pseudogap and (ii) that it is the unique doping dependence of the 
magnetic response which establishes the first numerical proof that 
the pseudogap is due to antiferromagnetic spin correlations in the 
under-doped regime.

\acknowledgments{
One of us (W. H.) would like to thank Z.-X. Shen, J. R. Schrieffer, 
D. J. Scalapino, E. Dagotto, A. Moreo, E. Arrigoni and R. Noack for 
instructive comments. Support by FORSUPRA II, BMBF (05 605 WWA 6), 
ERB CHRXCT940438 and computational facilities at the HLRZ J\"ulich 
and LRZ M\"unchen are acknowledged.}


\begin{references}
\bibitem{15a}{The average error level of all QMC-data was well below 0.5\%,
    except for $<n>$=0.93, $T$=0.25$t$, where the error level is higher due
    to sign problems.}
\bibitem{18a}{Because the integrated spectral weight suffers less from
    resolution problems of the Maximum-Entropy technique, we examined 
    in detail the temperature dependence 
    of the total spectral weight in the quasi-particle band at $(\pi,\pi)$. 
    This signal is calculated from the percentage of weight in this band 
    (which is above the chemical potential $\mu$ and well separated from 
    the incoherent background, see Fig. \ref{f02}(a)) with respect to the
    total intensity, $\int^{+\infty}_{-\infty} d\omega A(\vec k,\omega)=1$.
    At $T=0.5t$ and $\langle n \rangle=0.93$ we have $\approx 14\%$ 
    intensity, at $T=0.33t$ we get $\approx 9\%$ and, finally, less 
    than $5\%$ at $T=0.25t$. In ARPES a weak signal with less than roughly 
    $10\%$ intensity of the largest signal (located both in theory 
    and experiment at $(\pi/2,\pi/2)$) is not observable \cite{25}. 
    In a similiar way our QMC results in Fig. \ref{f02}(b) at $T=0.25t$ 
    can be interpreted: the resolution of the Maximum-Entropy technique 
    for the QMC error level at this temperature is such that peaks with 
    relative intensities less than about $10\%$ will not be resolved. 
    This is the criterion followed in this work to classify a result 
    as observable in the dynamical QMC method.}
\bibitem{21}{The correlation length $\xi$ is obtained from a fit of
    the QMC equal-time spin-spin correlation function to a form 
    $e^{-r/\xi}$.}
\end{references}
\end{document}